\documentclass[aps,prd,preprint,twoside,showpacs,superscriptaddress,nofootinbib,amsmath,amssymb]{revtex4-1}

\usepackage{MnSymbol}
\usepackage[utf8]{inputenc}
\usepackage{soulutf8}
\usepackage{graphicx}
\usepackage{hyperref}
\usepackage{epstopdf}
\usepackage{amsmath}
\usepackage{slashed}
\usepackage{xcolor}
\usepackage{mathtools}
\usepackage{array,multirow}
\usepackage{booktabs}
\usepackage{adjustbox} 
\usepackage{rotating}
\usepackage{blindtext}
\usepackage{physics}
\usepackage{caption}
\usepackage{subcaption}

\colorlet{darkred}{red!80!black}
\colorlet{darkgreen}{green!50!black}
\colorlet{darkblue}{blue!50!black}

\begin{document}

\title{Dynamical gluon effects in the three-dimensional structure of pion}

\author{Jiangshan Lan}
\email{jiangshanlan@impcas.ac.cn}
\affiliation{Institute of Modern Physics, Chinese Academy of Sciences, Lanzhou, Gansu 730000, China}
\affiliation{Advanced Energy Science and Technology Guangdong Laboratory, Huizhou, Guangdong 516000, China}
\affiliation{School of Nuclear Physics, University of Chinese Academy of Sciences, Beijing 100049, China}
\affiliation{CAS Key Laboratory of High Precision Nuclear Spectroscopy, Institute of Modern Physics, Chinese Academy of Sciences, Lanzhou 730000, China}

\author{Kaiyu Fu}
\email{kaiyufu94@gmail.com}
\affiliation{Institute of Modern Physics, Chinese Academy of Sciences, Lanzhou, Gansu 730000, China}

\author{Satvir Kaur}
\email{satvir@impcas.ac.cn}
\affiliation{Institute of Modern Physics, Chinese Academy of Sciences, Lanzhou, Gansu 730000, China}
\affiliation{School of Nuclear Physics, University of Chinese Academy of Sciences, Beijing 100049, China}
\affiliation{CAS Key Laboratory of High Precision Nuclear Spectroscopy, Institute of Modern Physics, Chinese Academy of Sciences, Lanzhou 730000, China}

\author{Zhimin Zhu}
\email{zhuzhimin@impcas.ac.cn}
\affiliation{Institute of Modern Physics, Chinese Academy of Sciences, Lanzhou, Gansu 730000, China}
\affiliation{School of Nuclear Physics, University of Chinese Academy of Sciences, Beijing 100049, China}
\affiliation{CAS Key Laboratory of High Precision Nuclear Spectroscopy, Institute of Modern Physics, Chinese Academy of Sciences, Lanzhou 730000, China}

\author{Chandan Mondal}
\email{mondal@impcas.ac.cn}
\affiliation{Institute of Modern Physics, Chinese Academy of Sciences, Lanzhou, Gansu 730000, China}
\affiliation{School of Nuclear Physics, University of Chinese Academy of Sciences, Beijing 100049, China}
\affiliation{CAS Key Laboratory of High Precision Nuclear Spectroscopy, Institute of Modern Physics, Chinese Academy of Sciences, Lanzhou 730000, China}

\author{Xingbo Zhao}
\email{xbzhao@impcas.ac.cn}
\affiliation{Institute of Modern Physics, Chinese Academy of Sciences, Lanzhou, Gansu 730000, China}
\affiliation{Advanced Energy Science and Technology Guangdong Laboratory, Huizhou, Guangdong 516000, China}
\affiliation{School of Nuclear Physics, University of Chinese Academy of Sciences, Beijing 100049, China}
\affiliation{CAS Key Laboratory of High Precision Nuclear Spectroscopy, Institute of Modern Physics, Chinese Academy of Sciences, Lanzhou 730000, China}

\author{James P. Vary}
\email{jvary@iastate.edu}
\affiliation{Department of Physics and Astronomy, Iowa State University, Ames, Iowa 50011, USA}

\collaboration{BLFQ Collaboration}

\begin{abstract} 
We investigate the internal structure of the pion, including the contributions from one dynamical gluon, using the basis light-front quantization (BLFQ) approach. By solving a light-front QCD Hamiltonian with a three-dimensional confining potential, we obtain the light-front wavefunctions (LFWFs) for both the quark-antiquark and quark-antiquark-gluon Fock sectors. These wavefunctions are then employed to compute the unpolarized generalized parton distributions (GPDs) and the transverse-momentum-dependent parton distributions (TMDs) of valence quarks and gluons. We also extract the transverse spatial distributions, providing the squared radii of quark and gluon densities in the impact-parameter space. This work contributes toward a three-dimensional understanding of the pion’s internal structure in both momentum and coordinate space.

\end{abstract}

\maketitle

\section{Introduction}
The pion, as the lightest meson and a Nambu-Goldstone boson of spontaneously broken chiral symmetry in quantum chromodynamics (QCD)~\cite{Nambu:1961tp}, plays a vital role in shaping the structure of matter in the universe.  It is not only essential for understanding the origin of hadron masses but also contributes to the stability of nuclear matter by acting as a mediating particle between nucleons. Probing its three-dimensional (3D) structure, particularly through generalized parton distributions (GPDs)~\cite{Diehl:2003ny,Ji:1998pc,Radyushkin:2000uy} and transverse momentum-dependent parton distributions (TMDs)~\cite{Barone:2001sp,Rogers:2015sqa,Angeles-Martinez:2015sea,Diehl:2015uka,Boussarie:2023izj}, offers essential insights into how relativistically moving quarks and gluons (partons) behave inside the pion in both position and momentum space. These studies also shed light on the orbital angular momentum (OAM) and spin contributions of the partons to the overall spin of the pion.

GPDs, defined in terms of the momentum fraction \(x\), skewness \(\xi\), and momentum transfer squared \(t\), encode rich information about the spatial and momentum structure of hadrons. Although not probabilistic in momentum space, their 2D Fourier transforms in the transverse plane (at \(\xi = 0\)) allow for a spatial interpretation of partons inside hadrons~\cite{Burkardt:2000za, Burkardt:2002hr}. Complementing this, TMDs offer a momentum space perspective in terms of ($x,{k}^2_\perp$), where ${\vec{k}}_\perp$ denotes the transverse momentum of the parton. Various limits of GPDs and TMDs connect them to parton distribution functions (PDFs), form factors, and the angular momentum contributions of quarks and gluons.

Considerable efforts have been made to investigate the internal structure of the pion using various theoretical frameworks. The leading-twist quark GPDs of the pion have been investigated using approaches such as lattice QCD~\cite{Chen:2019lcm,Lin:2023gxz,Ding:2024saz}, covariant approaches like Dyson-Schwinger equations (DSE) and Bethe-Salpeter equation (BSE)~\cite{Albino:2022gzs,Raya:2022eqa,Raya:2021zrz,Mezrag:2014jka,Chang:2015ela}, non-local chiral quark model~\cite{Son:2024uet} and light-front quark models~\cite{Choi:2002ic,deTeramond:2018ecg,Kaur:2018ewq,Kaur:2020vkq,Angulo:2022edl,Shi:2020pqe}
, revealing important spatial correlations of partons within the pion. However, only a few studies have been carried out for the gluon GPDs in pion~\cite{Broniowski:2022iip}.

Likewise, several studies have modeled pion TMDs to explore the intrinsic transverse motion of quarks and their spin-momentum correlations, often within phenomenological, covariant approaches or light-front frameworks~\cite{Signori:2013gra,Pasquini:2014ppa,Lorce:2016ugb,Broniowski:2017gfp,Shi:2018zqd,Ahmady:2019yvo,Kaur:2019jfa,Kaur:2020vkq,Shi:2020pqe}. Most analyses to date focus on the valence quark sector, with limited attention given to gluonic contributions and spin-related structures. A lattice QCD study of the Boer–Mulders effect in the pion has also been reported~\cite{Engelhardt:2015xja}. More recently, pion TMDs have been extracted from Drell–Yan data by the MAP collaboration~\cite{Cerutti:2022lmb}.

Lattice QCD~\cite{Wilson:1974sk} and DSE~\cite{Roberts:1994dr}, as first-principles approaches formulated in Euclidean space-time, do not offer direct access to parton distributions, which are defined on the light front. In contrast, basis light-front quantization (BLFQ)~\cite{Vary:2009gt}, developed in Minkowski space-time, enables direct computation of these distributions on the light front and holds the potential to solve QCD from first principles. While BLFQ is progressing toward a first-principles framework, particularly in the nucleon sector~\cite{Xu:2024sjt}, its application to mesons has thus far included interactions only from the lowest Fock sectors, namely the quark-antiquark (\(|q\bar{q}\rangle\)) and quark-antiquark-gluon (\(|q\bar{q}g\rangle\)) states~\cite{Lan:2021wok}. Our focus here will be on the dynamical gluon contributions to the properties of the pion.

This work builds on earlier BLFQ studies that investigated the pion PDFs, electromagnetic form factors, and higher-twist TMDs using both leading and next-to-leading Fock sectors~\cite{Lan:2019img,Lan:2021wok,Zhu:2023lst}. The BLFQ approach has also been applied to study the structure of other mesons, particularly the $\rho$ meson (a spin-$1$ meson)~\cite{Kaur:2024iwn} and, more recently, the kaon~\cite{Lan:2025fia}. In this work, we particularly focus on the leading-twist 3D structure of the pion through its unpolarized GPDs and TMDs, while also exploring its unpolarized gluon distributions.

Experimentally, GPDs appear naturally in the QCD framework for hard exclusive processes such as deeply virtual Compton scattering (DVCS)~\cite{Ji:1996nm,Marukyan:2015owj,Collins:1998be} and meson electroproduction (DVMP)~\cite{Amrath:2008vx}. In the case of pseudoscalar mesons, GPDs can be accessed experimentally through processes like the Sullivan process~\cite{Sullivan:1971kd}, where a virtual meson is probed by an off-shell photon, effectively enabling DVCS on a mesonic target. These exclusive reactions are of particular interest at several experimental facilities—such as Jefferson Lab~\cite{Accardi:2023chb}, the upcoming Electron-Ion Collider in the USA (EIC)~\cite{Arrington:2021biu,Accardi:2012qut} and China (EicC)~\cite{Anderle:2021wcy}, and J-PARC~\cite{Sawada:2016mao}—that aim to achieve wide kinematic coverage and high luminosity. The resulting data are expected to significantly improve our knowledge of meson GPDs and provide a more complete picture of how quarks and gluons are distributed within hadrons in both momentum and coordinate space.

On the other hand, TMDs are primarily accessed through semi-inclusive deep inelastic scattering (SIDIS)~\cite{Brodsky:2002cx}, where a hadron is detected alongside the scattered lepton, and through pion-induced Drell-Yan processes~\cite{Drell:1970wh,Drell:1970wh}. Additional channels like di-hadron and jet production~\cite{Angeles-Martinez:2015sea} also provide valuable access to TMDs.

The EICs are poised to play a leading role in TMD studies, with their broad kinematic coverage and high luminosity enabling precise measurements of key observables across multiple processes. These experimental efforts will deepen our understanding of parton dynamics, including spin-orbit interactions and orbital angular momenta, thereby contributing to the broader quest of mapping the full 3D structure of hadrons~\cite{AbdulKhalek:2021gbh}.
In parallel, our theoretical investigations of pion TMDs using the BLFQ approach aim to provide complementary insights and benchmark predictions that can guide and be tested against forthcoming EIC experiments. This synergy between theory and experiment is essential for advancing a complete understanding of the pion's internal dynamics.


\section{Basis Light-Front Quantization (BLFQ)}



The bound-state problem on the light front (LF) is formulated as an eigenvalue equation of the Hamiltonian:
\begin{equation}
P^- P^+|\Psi\rangle = M^2 |\Psi\rangle,
\end{equation}
where $P^\pm = P^0 \pm P^3$ are the LF energy ($P^-$) and longitudinal momentum ($P^+$), and $M^2$ is the invariant mass squared of the state. At the fixed LF time $x^+ = t + z$, the meson state can be expanded in terms of Fock components consisting of quarks ($q$), antiquarks ($\bar{q}$), and gluons ($g$)~\cite{Brodsky:1997de},
\begin{align}\label{Eq1}
|\Psi\rangle = \psi_{\lbrace q\bar{q} \rbrace}|q\bar{q}\rangle + \psi_{\lbrace q\bar{q}g \rbrace}|q\bar{q}g\rangle + \dots,
\end{align}
where the light-front wave functions (LFWFs) $\psi_{\lbrace \dots \rbrace}$ denote the probability amplitudes for various partonic configurations.

At the initial scale, we include the $|q\bar{q}\rangle$ and $|q\bar{q}g\rangle$ sectors, and use the LF Hamiltonian
\begin{equation}
P^- = P^-_{\mathrm{QCD}} + P^-_{\mathrm{C}},
\end{equation}
where $P^-_{\mathrm{QCD}}$ is the QCD Hamiltonian and $P^-_{\mathrm{C}}$ introduces a model for confinement. In the light-front gauge $A^+=0$, and with one dynamical gluon~\cite{Brodsky:1997de}, the QCD Hamiltonian is given by
\begin{align}
P^-_{\mathrm{QCD}} &= \int \mathrm{d}x^- \mathrm{d}^2\vec{x}^\perp  \Bigg\{
\frac{1}{2} \bar{\psi} \gamma^+ \frac{m_0^2 + (i\partial^\perp)^2}{i\partial^+} \psi 
+ \frac{1}{2} A_a^i \left[m_g^2 + (i\partial^\perp)^2\right] A^i_a \nonumber \\
&\quad + g_s \bar{\psi} \gamma_\mu T^a A_a^\mu \psi 
+ \frac{1}{2} g_s^2 \bar{\psi} \gamma^+ T^a \psi \frac{1}{(i\partial^+)^2} \bar{\psi} \gamma^+ T^a \psi 
\Bigg\}, \label{eqn:PQCD}
\end{align}
where $\psi$ and $A^\mu$ are the quark and gluon fields, respectively. 
$T^a$ are the $SU(3)$ generators in the adjoint representation, and $\gamma^+ = \gamma^0 + \gamma^3$.

The first two terms in Eq.~\eqref{eqn:PQCD} describe the kinetic energies of quarks and gluons, with $m_0$ and $m_g$ being their bare masses. While gluons are massless in QCD, we allow a phenomenological gluon mass $m_g$ to model the neglected contributions from QCD that we expect to arise when including higher Fock Sectors. This is motivated by the Schwinger mechanism, where nonperturbative gluon self-energy corrections generate a gauge-invariant dressed gluon whose propagator behaves as if it has an effective mass~\cite{Cornwall:1981zr, Aguilar:2007nf, Binosi:2014aea}. In our work, $m^2_g$ serves as a phenomenological parameter consistent with infrared regularization in Hamiltonian approaches and effective models. The remaining terms describe interaction vertices governed by the coupling constant $g_s$.

Following the renormalization approach developed for positronium in truncated Fock sectors~\cite{Zhao:2014hpa,Zhao:2020kuf}, we introduce a mass counterterm $\delta m_q = m_0 - m_q$, where $m_q$ is the renormalized quark mass. An independent quark mass $m_f$ is introduced in the vertex interaction, as motivated in Ref.~\cite{Glazek:1992aq}.

For the confinement in the leading $|q\bar{q}\rangle$ sector, we adopt the following potential~\cite{Li:2015zda}:
\begin{align}
P^-_{\mathrm{C}} P^+ &= \kappa^4 \left\{ x(1-x) \vec{r}_\perp^2 
- \frac{\partial_x [x(1-x) \partial_x]}{(m_q + m_{\bar{q}})^2} \right\}, \label{eqn:PC}
\end{align}
where $\kappa$ is the confinement strength. The transverse part reproduces the LF holographic potential, with the light-front holographic variable $\vec{r}_\perp = \sqrt{x(1-x)} (\vec{r}_{\perp q} - \vec{r}_{\perp \bar{q}})$~\cite{Brodsky:2014yha}. This potential yields a symmetric 3D confinement in the nonrelativistic limit and has been applied to both mesons and baryons~\cite{Li:2021jqb,DeTeramond:2021yyi,Li:2015zda,Jia:2018ary,Lan:2019vui,Lan:2019rba,Tang:2018myz,Tang:2019gvn,Lan:2019img,Qian:2020utg,Mondal:2019jdg}. In the $|q\bar{q}g\rangle$ sector, confinement is accommodated by the truncation of the basis functions used in the BLFQ framework as well as the finite gluon mass $m_g$.

In BLFQ, each Fock particle is described by a longitudinal plane wave $e^{-i p^+ x^-/2}$ and a transverse 2D harmonic oscillator (2D-HO) wavefunction $\Phi_{nm}(\vec{p}_\perp;b)$, with the scale parameter $b$~\cite{Zhao:2014xaa}. The longitudinal coordinate is confined in a box of size $2L$, with antiperiodic (periodic) boundary conditions for fermions (bosons). The longitudinal momenta are quantized as $p^+ = 2\pi k/L$, with $k = \frac{1}{2}, \frac{3}{2}, \dots$ for fermions and $k = 1, 2, \dots$ for bosons. The boson's zero mode is omitted.

Total longitudinal momentum is rescaled as $P^+ = \sum_i p^+_i = \frac{2\pi}{L} K$, where $K = \sum_i k_i$ defines the dimensionless total longitudinal momentum. The momentum fraction for the $i$-th parton is then $x_i = k_i / K$. Each parton basis state is labeled by quantum numbers $\bar{\alpha} = \{k, n, m, \lambda\}$, where $k$ is the longitudinal momentum quantum number mentioned above, $n$ and $m$ are the radial and angular quantum numbers of the 2D-HO, and $\lambda$ is the helicity.

In cases where multiple color singlet states are possible, additional labels are needed to distinguish them. However, since we consider only the \( |q\bar{q}\rangle \) and \( |q\bar{q}g\rangle \) Fock sectors in the present work, each of which contains a single color singlet state, we omit the color singlet label.
The total angular momentum projection of a many-body basis state is
\begin{equation}
M_J = \sum_i (m_i + \lambda_i).
\end{equation}

We introduce truncation parameters $K$ and $N_{\mathrm{max}}$ to restrict the basis size. The transverse truncation condition is $\sum_i (2n_i + |m_i| + 1) \le N_{\mathrm{max}}$, which ensures the factorization of the transverse center-of-mass motion~\cite{Wiecki:2014ola}, and introduces a natural UV cutoff $\Lambda_{\mathrm{UV}} \sim b \sqrt{N_{\mathrm{max}}}$, and an IR cutoff $\Lambda_{\mathrm{IR}} \sim b /\sqrt{N_{\mathrm{max}}}$ in the momentum space~\cite{Zhao:2014xaa}.

The LFWFs in momentum space are schematically expressed as
\begin{align}
\Psi^{\mathcal{N},\, M_J}_{\{x_i, \vec{k}_{\perp i}, \lambda_i\}} 
= \sum_{\{n_i, m_i\}} \psi^{\mathcal{N}}(\{\bar{\alpha}_i\}) 
\prod_{i=1}^{\mathcal{N}} \Phi_{n_i m_i}(\vec{k}_{\perp i}, b), \label{eqn:wf}
\end{align}
where $\psi^{\mathcal{N}=2}$ and $\psi^{\mathcal{N}=3}$ are the wavefunction components in the $|q\bar{q}\rangle$ and $|q\bar{q}g\rangle$ sectors, respectively, obtained from diagonalizing the full Hamiltonian. The functions {$\Phi_{nm}(\vec{k}_\perp, b)$} are the 2D-HO basis functions in momentum space.

The parameters of the model are fixed by fitting the masses of unflavored light mesons, as discussed in Ref.~\cite{Lan:2021wok}. Table~\ref{table:parameters} summarizes the parameter values used in previous studies of the pion, including its electromagnetic form factors, quark and gluon PDFs, and the differential cross section for $J/\psi$ production in pion-nucleus collisions~\cite{Lan:2021wok}. In the current work, we use the same parameters to extend our investigation of the pion structure to its 3D imaging through GPDs and TMDs.

\begin{table}[h]
  \caption{The list of the model parameters \cite{Lan:2021wok}.  All quantities are in units of [GeV] except $g_s$. }
    \vspace{0.15cm}
      \label{table:parameters}
  \centering
  \begin{tabular}{cccccc}
  \hline\hline 
         $m_q$ & $m_g$&$b$&$\kappa$ &${m}_f$ &$g_s$ \\        \hline 
                     0.39 & 0.60 &0.29 &0.65&5.69&1.92 \\
   \hline \hline
  \end{tabular}
\end{table}


\section{Transverse Momentum-dependent Distributions (TMDs)}
Being three-dimensional distribution functions, TMDs provide probabilistic information about observing a parton in momentum space ($x, {k}^2_\perp$). At uniform light-front time $z^+ = 0$, the unpolarized quark TMD is defined in terms of a correlator connecting two quark fields at different positions as~\cite{Goeke:2005hb,Bacchetta:2006tn,Meissner:2008ay},
\begin{align}
    f_{1,\pi}^q(x, {k}^2_\perp) 
    = \frac{1}{2}\int \frac{\mathrm{d}z^- \mathrm{d}^2 \vec{ z}_\perp}{2(2\pi)^3} \, 
    e^{i k\cdot z}
    \left\langle \pi(p) \Big| \bar{\psi}\left(-\frac{z}{2}\right) \gamma^+ \psi\left(\frac{z}{2}\right) \Big| \pi(p) \right\rangle_{z^+ = 0},
    \label{eqn:tmd1}
\end{align}
and the gluon TMD is defined as~\cite{Mulders:2000sh,Meissner:2007rx},
\begin{align}
f_{1,\pi}^g(x, {k}^2_\perp) 
= \frac{1}{x P^+} \int \frac{\mathrm{d}z^- \mathrm{d}^2 \vec{ z}_\perp}{2(2\pi)^3} \, 
e^{i k\cdot z}
\left\langle \pi(p) \Big| G^{+i}\left(-\frac{z}{2}\right) G^{+i}\left(\frac{z}{2}\right) \Big| \pi(p) \right\rangle_{z^+ = 0},
\label{eqn:tmd2}
\end{align}
where $G^{\mu\nu}$ is the gluon field strength tensor and a summation over $i=1,\,2$ is implied.
Here, we omit the gauge links in both correlators of Eqs.~(\ref{eqn:tmd1}) and (\ref{eqn:tmd2}) for simplicity---that is we adopt unit matrix approximations for the gauge links in this work.


In terms of the overlap of LFWFs, the quark and gluon TMDs are given by:
\begin{align}
&f_{1,\pi}^q(x, {k}^2_\perp) = \sum_{\mathcal{N},\lambda_i} \int_{\mathcal{N}} \left|\Psi^{\mathcal{N},M_J=0 }_{\{x_i,\vec{k}_{\perp i},\lambda_i\}} \right|^2 \delta(x-x_q)\delta^2( {\vec{k}}_\perp- {\vec{k}}_{\perp q}),
\end{align}
and
\begin{align}
&f_{1,\pi}^g(x, {k}^2_\perp) = \sum_{\mathcal{N},\lambda_i} \int_{\mathcal{N}} \left|\Psi^{\mathcal{N},M_J=0 }_{\{x_i,\vec{k}_{\perp i},\lambda_i\}}  \right|^2\delta(x-x_g)\delta^2( {\vec{k}}_\perp- {\vec{k}}_{\perp g}),
\label{eqn:tmdwf}
\end{align}
respectively,
where the integration measure is defined as
\begin{align}
    \int_{\mathcal{N}} \equiv \prod_{i=1}^{\mathcal{N}} \int \left[ \frac{\mathrm{d}x_i \, \mathrm{d}^2\vec{k}_{\perp i}}{16\pi^3} \right] 16\pi^3 \delta\left(1 - \sum_j x_j\right) \delta^{2}\left(\sum_j \vec{k}_{\perp j}\right). \label{eq:int}
\end{align}
Here, the index \( i = q, \bar{q}, g \) runs over the quark, antiquark, and gluon. For the quark TMD \( f^q_{1,\pi} \), contributions from both the leading (\( \mathcal{N} = 2 \)) and next-to-leading (\( \mathcal{N} = 3 \)) Fock sectors are included. In contrast, the gluon TMD \( f^g_{1,\pi} \) arises only from the next-to-leading sector (\( \mathcal{N} = 3 \)).


\begin{figure}[hbt!]
     \centering
     \begin{subfigure}[b]{0.47\textwidth}
         \centering
         \includegraphics[width=\textwidth]{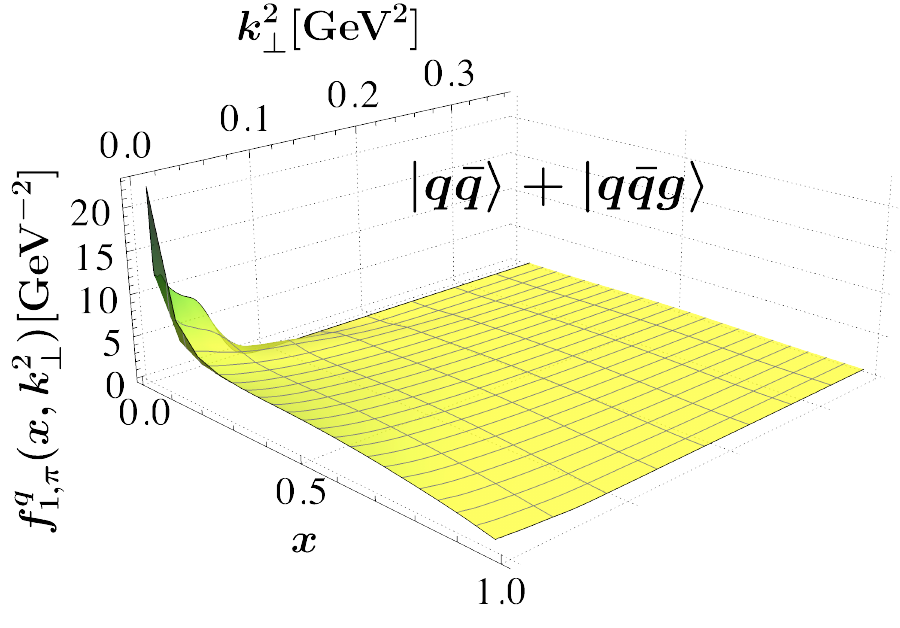}
         \caption{}
         \label{fig:tmd-total}
     \end{subfigure}
    \\
     \begin{subfigure}[b]{0.47\textwidth}
         \centering
         \includegraphics[width=\textwidth]{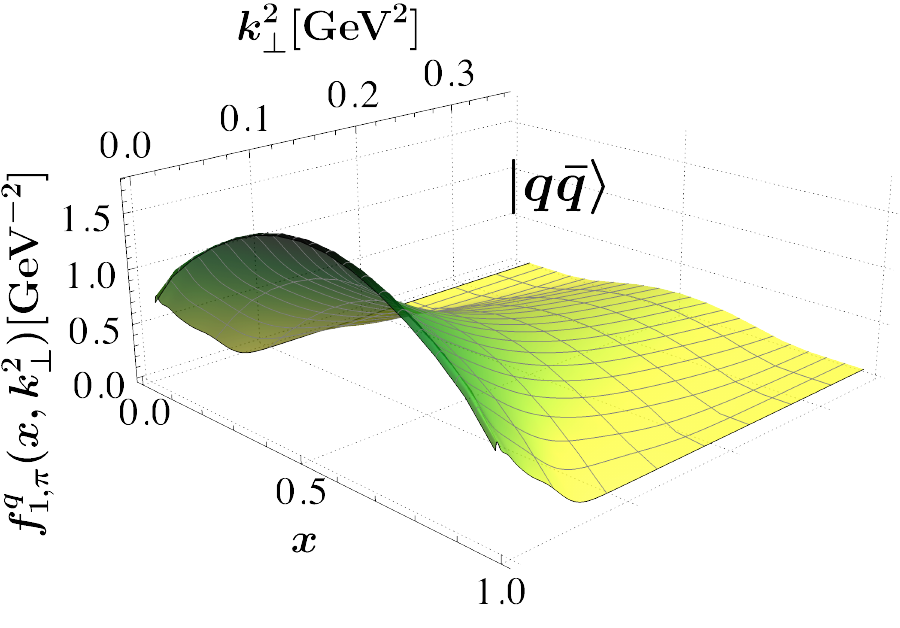}
         \caption{}
         \label{fig:tmd-q-Fock1}
     \end{subfigure}
     \begin{subfigure}[b]{0.47\textwidth}
         \centering
         \includegraphics[width=\textwidth]{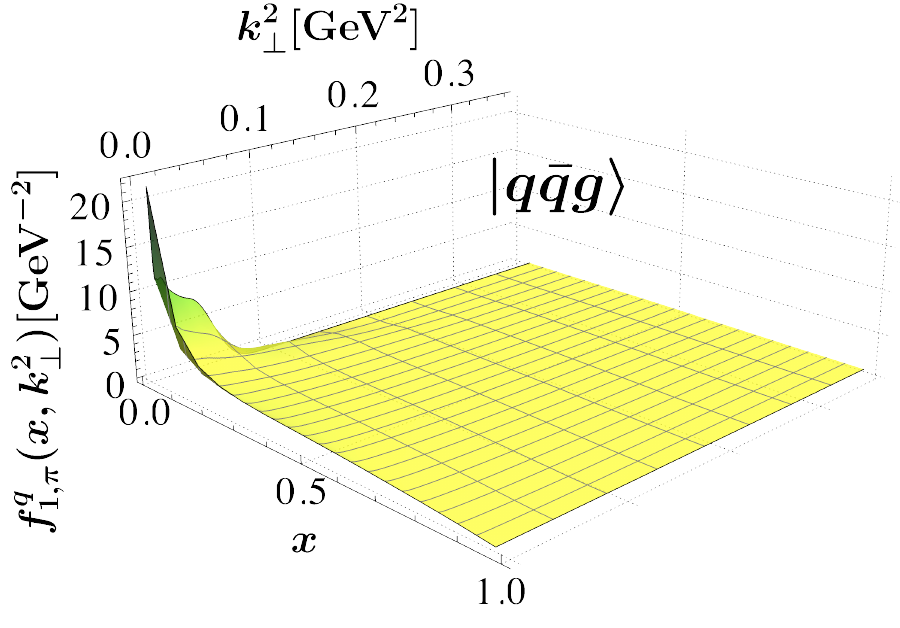}
         \caption{}
         \label{fig:tmd-q-Fock2}
     \end{subfigure}
             \caption{
             The unpolarized TMD of quark in the Fock sectors (a) $|q\bar{q}\rangle+|q\bar{q}g \rangle$, (b) $|q\bar{q}\rangle$, and (c) $|q \bar{q}g\rangle$ with respect to $x$ and $k^2_\perp$.}
        \label{fig:tmds}
\end{figure}

\begin{figure}
    \centering
    \includegraphics[width=0.5\linewidth]{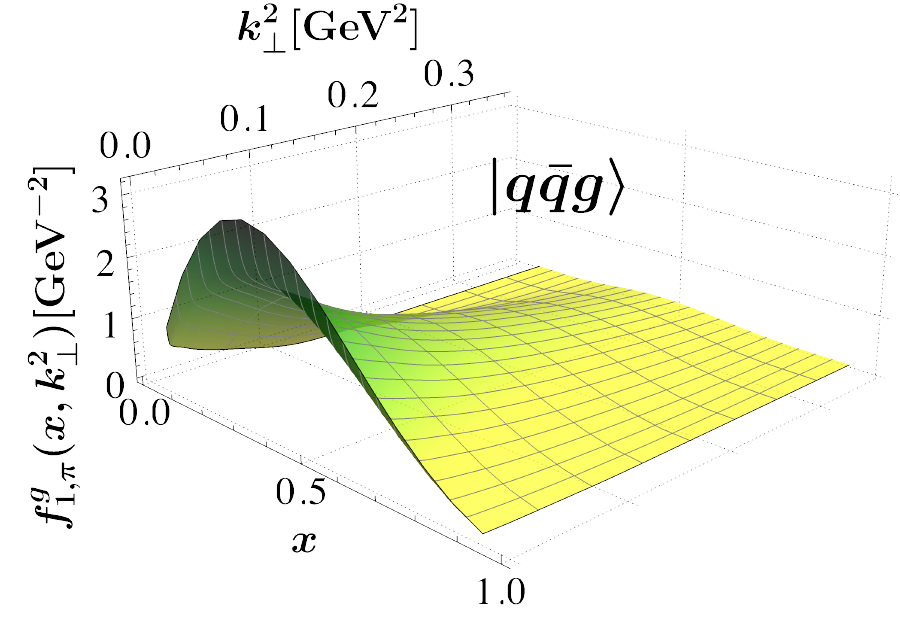}
    \caption{The unpolarized gluon TMD with respect to $x$ and $k^2_\perp$.}
    \label{fig:tmd-g-Fock2}
\end{figure}



In Fig.~\ref{fig:tmds}, we present the 3D profiles of the unpolarized quark and gluon TMDs, $f_1(x,k^2_\perp)$, in the pion, as functions of the longitudinal momentum fraction $x$ and the square of the transverse momentum, $k^2_\perp$, carried by the parton. The total TMDs, incorporating contributions from both the leading and next-to-leading Fock sectors, are displayed in Fig.~\ref{fig:tmd-total}, while the individual contributions from the leading ($|q\bar{q}\rangle$) and next-to-leading ($|q\bar{q}g\rangle$) sectors are shown separately in Fig.~\ref{fig:tmd-q-Fock1} and Fig.~\ref{fig:tmd-q-Fock2}, respectively.

As expected, the quark TMD shows a peak at low \( x \), highlighting the impact of the next-to-leading Fock sector that includes a dynamical gluon, as shown in Fig.~\ref{fig:tmd-q-Fock2}. 
In contrast, the leading (valence) Fock sector contributes predominantly at intermediate values of \( x \), with a broader distribution in transverse momentum \( k_\perp^2 \), as seen in Fig.~\ref{fig:tmd-q-Fock1}. 
Additionally, the TMD from the valence sector falls more slowly at large \( k_\perp^2 \) compared to the next-to-leading sector. 
This slower fall-off indicates that valence quarks are more likely to carry higher transverse momenta, suggesting stronger intrinsic transverse motion in the absence of explicit gluon degrees of freedom. 
Such momentum-space behavior has been observed in earlier phenomenological studies~\cite{Kaur:2020vkq,Shi:2020pqe,Pasquini:2014ppa,Lorce:2016ugb,Shi:2018zqd,Ahmady:2019yvo,Kaur:2019jfa}.

The inclusion of the gluon component introduces significant modifications to the TMD, particularly in the low-$x$ region, and moves the description closer to a more complete momentum-space picture of the pion. 
The combined TMD shown in Fig.~\ref{fig:tmd-total} begins to exhibit features reminiscent of QCD evolution, that are absent in a purely valence-based framework but emerge naturally when higher Fock components are taken into account.

In Fig.~\ref{fig:tmd-g-Fock2}, we show the gluon TMD in the pion, extracted from the \( q\bar{q}g \) Fock sector. The probability of finding an unpolarized gluon carrying a momentum fraction \( x \) in a pion peaks around \( x \approx 0.33 \). The distribution is narrower in \( x \) compared to the quark TMD from the valence Fock sector, indicating a more localized gluon momentum structure along the longitudinal direction. 
We also observe that the probability of finding a gluon is larger than that of finding a quark in the valence Fock sector at our model scale. 
Moreover, the overall shape of the gluon TMD is consistent with an effective massive gluon as it vanishes at both endpoints \( x \to 0 \) and \( x \to 1 \), and is concentrated in the intermediate-\( x \) region.





\subsection{The ${k}_\perp$-moment of pion TMDs}


To compare our results with available theoretical predictions, we calculate the transverse momentum moments of pion unpolarized TMD $f_1(x,k^2_\perp)$. The moments are defined as
\begin{align}
    \langle {k}_\perp^n \rangle_i(x) = \frac{\int \mathrm{d}^2{\vec{k}_\perp} {k}_\perp^n f_{1,\pi}^i(x,k^2_\perp)}{\int \mathrm{d}^2{\vec{k}_\perp} f_{1,\pi}^i(x,k^2_\perp)},
\end{align}
where $n$ denotes the order of the moment.

\begin{figure}
    \centering
    \includegraphics[width=0.5\linewidth]{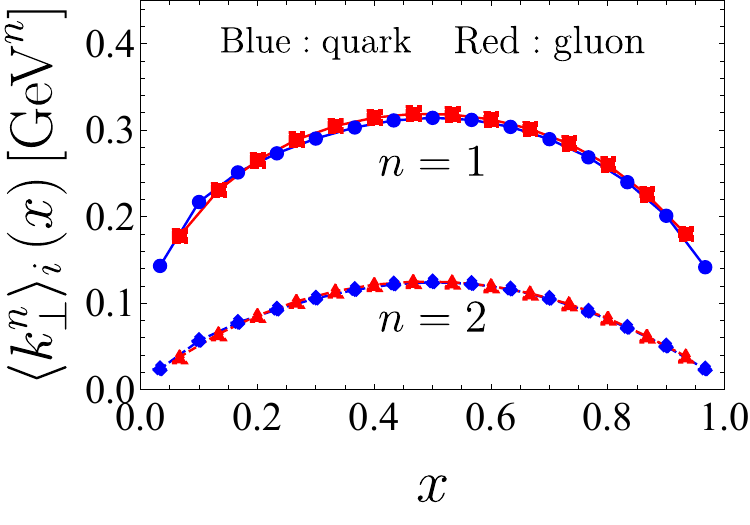}
    \caption{First and second ${k}_\perp$-moments of the quark and gluon TMDs in the pion as functions of the longitudinal momentum fraction $x$.}
    \label{fig:k-moment}
\end{figure}

Figure~\ref{fig:k-moment} presents the first and second ${k}_\perp$-moments of the quark and gluon TMDs in the pion as functions of $x$. The results show that the ${k}_\perp$-moments of quarks and gluons are very close to each other across the entire range of the momentum fraction $x$.

Further, in Table~\ref{Table:moments}, we present our predictions for the $x$-integrated first and second ${k}_\perp$-moments of the quark and the gluon using our BLFQ approach, and compare them with available results from the the LF constituent model (LFCM)~\cite{Lorce:2016ugb}, LF holographic QCD model (LFHM)~\cite{Puhan:2023ekt}, and LF quark model (LFQM)~\cite{Puhan:2023ekt}.

\begin{table}[h]
  \caption{
 The first moment $\langle {k}_\perp \rangle$ (in GeV) and the second moment $\langle k_\perp^2 \rangle$ (in GeV$^2$) of the quark and gluon TMDs in the pion are presented and compared with the corresponding predictions from the LF
constituent model (LFCM)~\cite{Lorce:2016ugb}, LF holographic QCD model (LFHM)~\cite{Puhan:2023ekt}, and LF quark model (LFQM)~\cite{Puhan:2023ekt}. 
  }
    \vspace{0.15cm}
  \centering
  \begin{tabular}{c|cccc}
  \hline\hline 
          & $\langle {k}_\perp\rangle_q$&$\langle k_\perp^2\rangle_q$&$\langle {k}_\perp\rangle_g$ &$\langle k^2_\perp\rangle_g$ \\        \hline 
     This work & 0.26 & 0.087 & 0.25 &  0.086\\
     LFCM~\cite{Lorce:2016ugb} & 0.28 & 0.100 & - & - \\
     LFHM~\cite{Puhan:2023ekt} & 0.24 & 0.073 & - & - \\
     LFQM~\cite{Puhan:2023ekt} & 0.22 & 0.068 & - & - \\
   \hline \hline
  \end{tabular}
  \label{Table:moments}
\end{table}

\section{Generalized Parton Distributions (GPDs)}

In addition to TMDs, GPDs also offer a 3D view of hadron structure, specifically revealing its spatial tomography. The unpolarized GPD $H(x, \xi, t)$ for pseudoscalar mesons
is defined through light-front bilocal correlation functions of the vector current~\cite{Diehl:2003ny, Meissner:2008ay,Zhang:2021shm,Chavez:2021llq}. 
This applies to both quark and gluon contributions, which are given as
\begin{align}
&H_\pi^q(x,\xi=0,-t)= \int \frac{\mathrm{d}z^-}{2(2\pi)}e^{ixP^+z^-}\left\langle\pi(p')\Big|\bar{q}\left(-\frac{z}{2}\right)\gamma^+q\left(\frac{z}{2}\right)\Big|\pi(p)\right\rangle_{z^+=0,z_\perp=0},
\end{align}
%
%
and
\begin{align}
&H_\pi^g(x,\xi=0,-t)=\frac{1}{xP^+} \int \frac{\mathrm{d}z^-}{(2\pi)}e^{ixP^+z^-}\left\langle\pi(p')\Big|G^{+i}\left(-\frac{z}{2}\right)G^{+i}\left(\frac{z}{2}\right)\Big|\pi(p)\right\rangle_{z^+=0,z_\perp=0}, 
\label{eqn:gpd}
\end{align}
where \( p \) and \( p' \) are the incoming and outgoing momenta of the meson, respectively, and \( \Delta = p' - p \) is the momentum transfer, with \( t = -\vec{\Delta}_\perp^2 \). In this work, we focus on the unpolarized GPDs at zero skewness (\( \xi = 0 \)) and in the Dokshitzer-Gribov-Lipatov-Altarelli-Parisi (DGLAP) region, \( \xi < x < 1 \).


The overlap form of unpolarized quark and gluon GPD in terms of LFWFs is defined as
\begin{align}
&H_\pi^q(x,0,-t)=\sum_{\mathcal{N},\lambda_i} \int_{\mathcal{N}} \Psi^{\mathcal{N},M_J=0 *}_{\{x_i,\vec{k}^\prime_{\perp i},\lambda_i\}} \Psi^{\mathcal{N},M_J=0}_{\{x_i,\vec{k}_{\perp i},\lambda_i\}} \delta(x-x_q),
\end{align}
and
\begin{align}
&H_\pi^g(x,0,-t)= \sum_{\mathcal{N},\lambda_i} \int_{\mathcal{N}} \Psi^{\mathcal{N},M_J=0 *}_{\{x_i,\vec{k}^\prime_{\perp i},\lambda_i\}} \Psi^{\mathcal{N},M_J=0}_{\{x_i,\vec{k}_{\perp i},\lambda_i\}} \delta(x-x_g),
\label{eqn:gpdwf}
\end{align}
where the integration measure is defined in Eq.~(\ref{eq:int}).
The shifted transverse momenta in the final-state wavefunctions are defined as
\[
\vec{k}_{\perp i}^\prime = 
\begin{cases}
\vec{k}_{\perp i} + (1 - x_i)\vec{\Delta}_\perp, & \text{for the struck parton}, \\
\vec{k}_{\perp i} - x_i\vec{\Delta}_\perp, & \text{for spectator partons}.
\end{cases}
\]

In the forward limit \( t = 0 \), where the initial and final states are identical (\( p = p' \)), the quark GPD reduces to the usual quark PDF, \( H_\pi^q(x,0,0) = q(x) \), while the gluon GPD reduces to the gluon momentum distribution, \( H_\pi^g(x,0,0) = g(x) \)\footnote{Our gluon distribution differs by a factor of $x$ from those of in Refs.~\cite{Diehl:2003ny,Kaur:2025gyr}: $xH^g_\pi(x)|_{\rm here}=H^g_\pi(x)|_{\text{\cite{Diehl:2003ny,Kaur:2025gyr}}}$.}.
Here, \( q(x) \) and \( g(x) \) denote the quark and gluon PDFs in the pion, respectively. 

\begin{figure}[hbt!]
     \centering
     \begin{subfigure}[b]{0.47\textwidth}
         \centering
         \includegraphics[width=\textwidth]{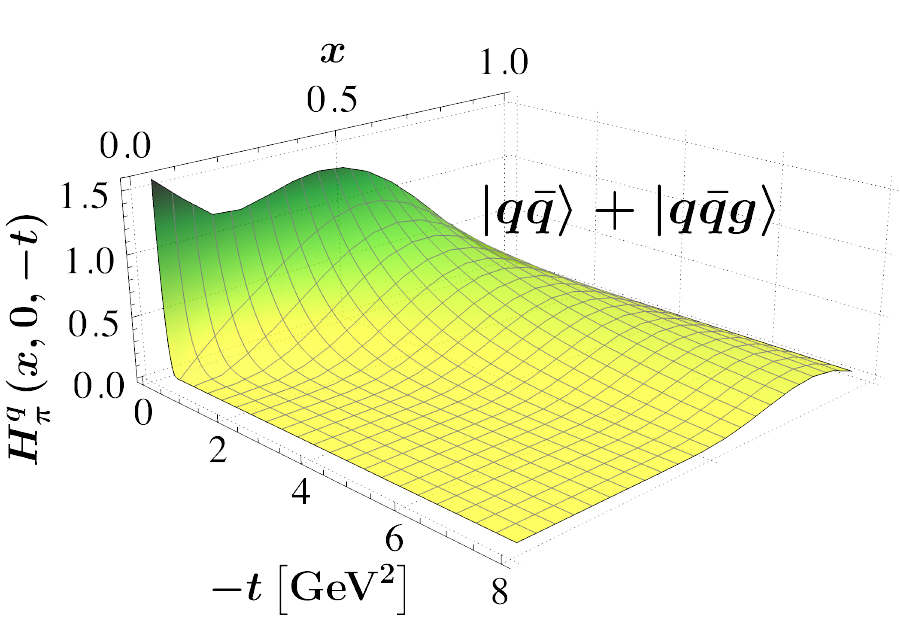}
         \caption{}
         \label{fig:gpd-total}
     \end{subfigure} 
     \\
     \begin{subfigure}[b]{0.47\textwidth}
         \centering
         \includegraphics[width=\textwidth]{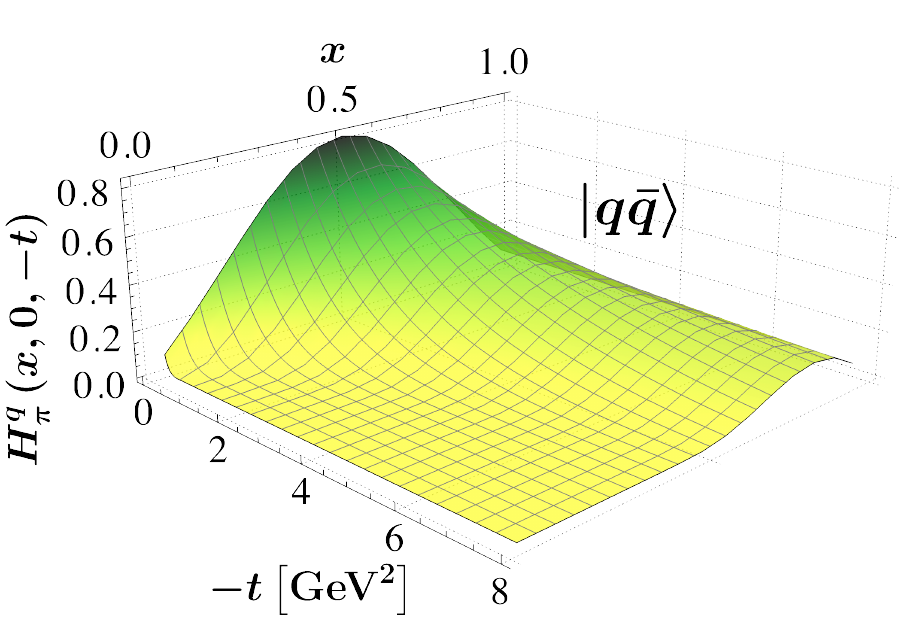}
         \caption{}
         \label{fig:gpd-q-Fock1}
     \end{subfigure}
     \hfill
     \begin{subfigure}[b]{0.47\textwidth}
         \centering
         \includegraphics[width=\textwidth]{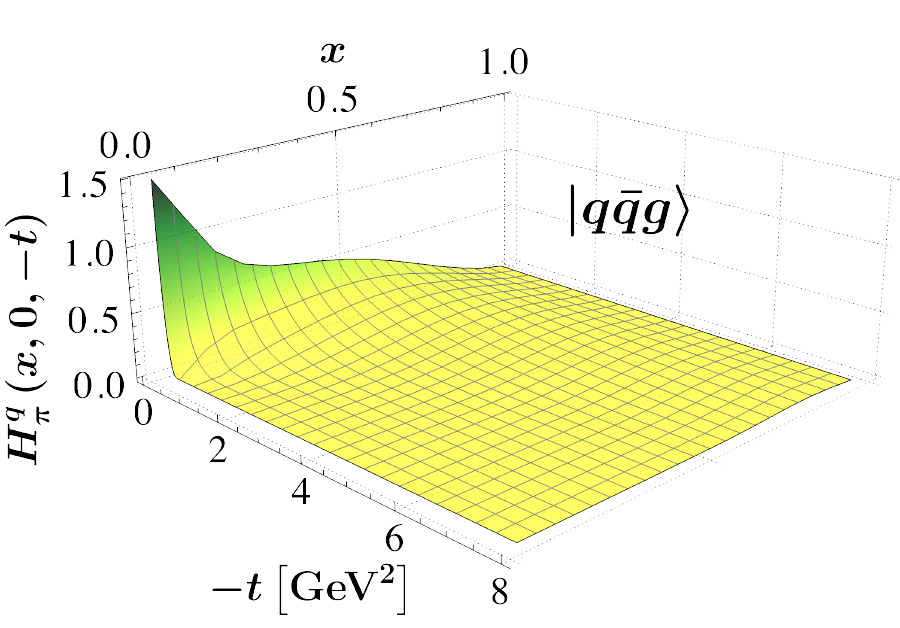}
         \caption{}
         \label{fig:gpd-q-Fock2}
     \end{subfigure}
             \caption{
             The unpolarized GPD of quark in the Fock sectors (a) $|q\bar{q}\rangle+|q\bar{q}g \rangle$, (b) $|q\bar{q}\rangle$, and (c) $|q \bar{q}g\rangle$ with respect to $x$ and $-t$.}
        \label{fig:gpds}
\end{figure}

\begin{figure}
    \centering
    \includegraphics[width=0.5\linewidth]{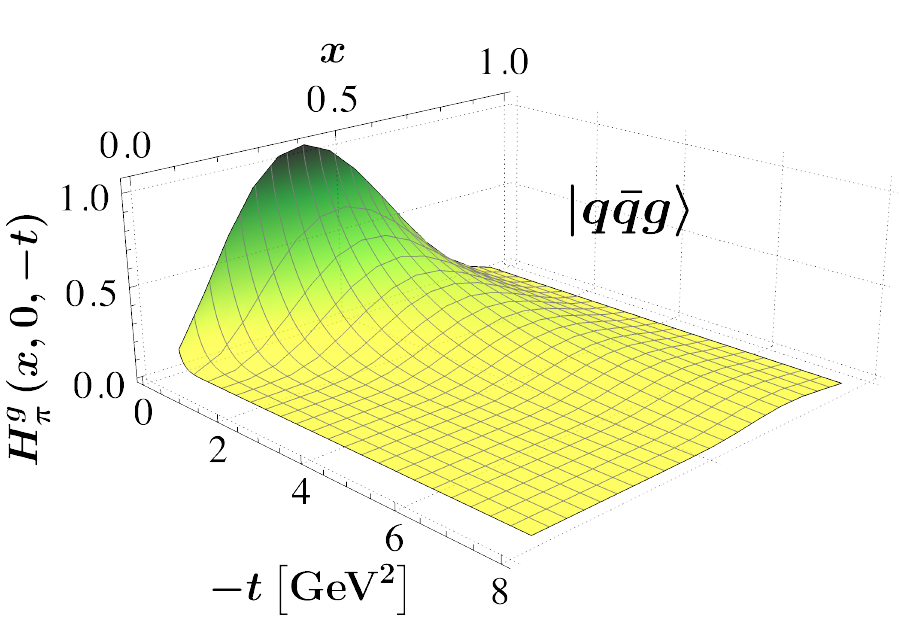}
    \caption{
    The unpolarized gluon GPD with respect to $x$ and $-t$.}
    \label{fig:gpd-g-Fock2}
\end{figure}


Using the computed LFWFs within the BLFQ framework, we evaluate the unpolarized GPD of the pion and present the results in Figs.~\ref{fig:gpds} and \ref{fig:gpd-g-Fock2}. These GPDs encode the correlated distributions of partons in both longitudinal momentum and transverse spatial degrees of freedom, providing a multidimensional view of the internal structure of the pion.

Figure~\ref{fig:gpd-total} shows the total quark GPD, highlighting the effects of incorporating a dynamical gluon. Similar to the TMDs, the impact from gluons is most significant in the small-\( x \) region, particularly at small $t$, as clearly seen in Fig.~\ref{fig:gpd-q-Fock2}. This enhancement at low \( x \) can be interpreted as the onset of gluon contributions, which are naturally incorporated through the inclusion of higher Fock sectors in our framework.

In the valence quark Fock sector (\( q\bar{q} \)), the GPD exhibits features consistent with previous model and phenomenological studies. As the momentum transfer \( -t \) increases, the distributions become increasingly skewed toward larger values of \( x \). This behavior reflects the tendency of high-momentum-transfer interactions to probe partons carrying larger fractions of the pion’s longitudinal momentum. Such \( t \)-dependent evolution of the GPDs has been reported across various studies involving mesons~\cite{Kaur:2018ewq,Adhikari:2018umb,Adhikari:2021jrh,deTeramond:2018ecg,Albino:2022gzs} and nucleons~\cite{Chakrabarti:2013gra,Mondal:2015uha,Xu:2021wwj}, indicating a largely model-independent trend.

The gluon GPD, shown in Fig.~\ref{fig:gpd-g-Fock2}, displays a narrower distribution in \( x \) compared to the quark GPD, and a relatively higher peak. The narrower distribution is a consequence of the larger effective mass assigned to the gluon in our model, which suppresses contributions from partons with small and large \( x \), thereby concentrating the strength in the intermediate-\( x \) region. Additionally, the gluon GPD exhibits a more rapid fall-off with increasing \( -t \) compared to the quark GPD, suggesting that the gluonic spatial distribution is more localized in the coordinate space. Overall, the behavior of our pion's gluon GPD is largely consistent with the gluon distribution obtained in a phenomenological model for the pion~\cite{Kaur:2025gyr}, as well as with the gluon GPDs in the proton previously computed within our BLFQ framework~\cite{Lin:2023ezw,Zhang:2025nll,Lin:2024ijo}.






  \begin{figure}
    \begin{center}
    \includegraphics[width=0.326\linewidth]{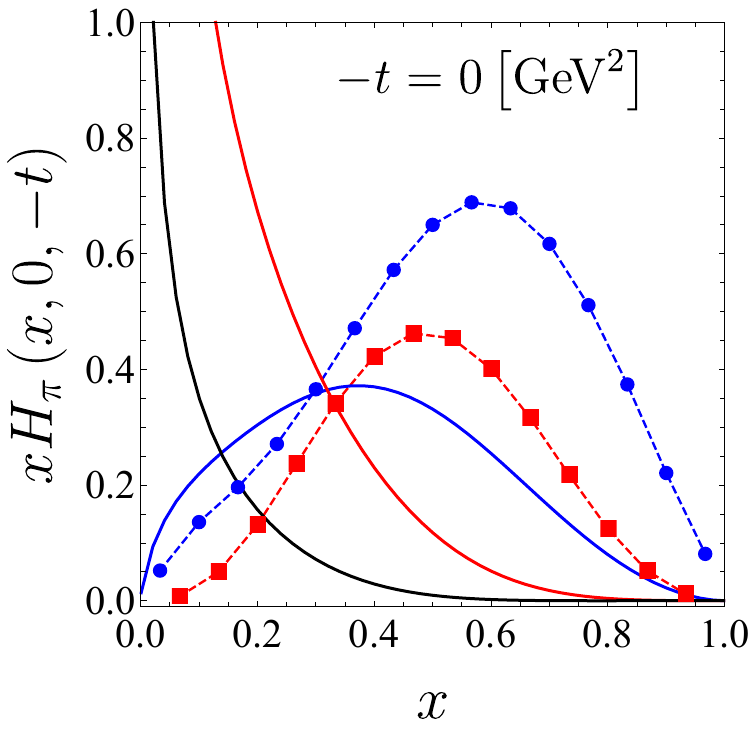}
    \includegraphics[width=0.326\linewidth]{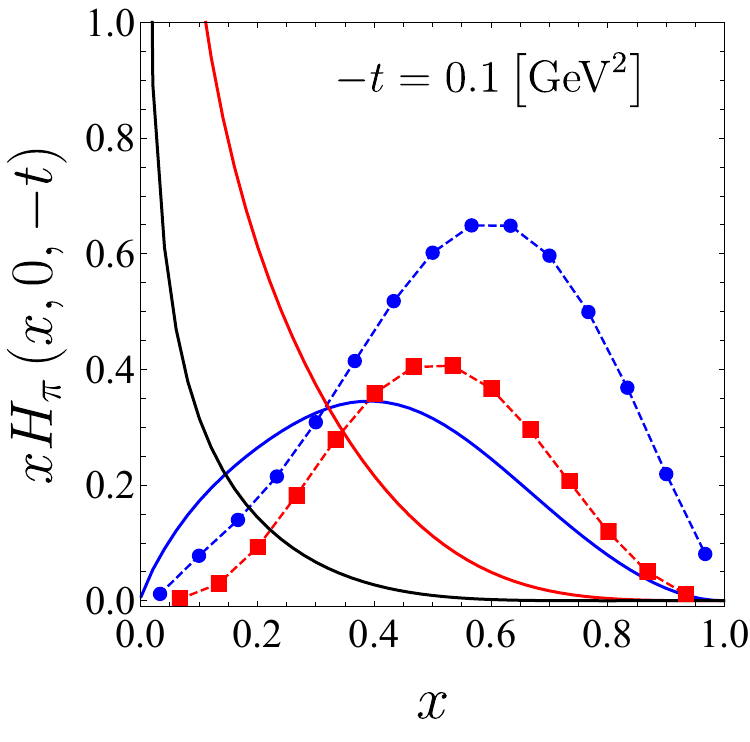}
    \includegraphics[width=0.326\linewidth]{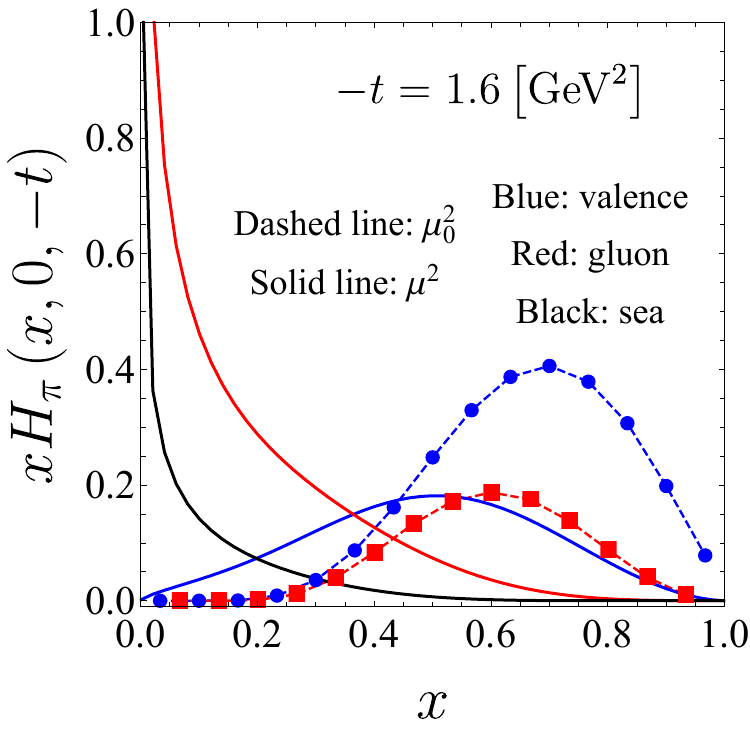}
    \caption{
    The $x$ multiplied pion GPDs at different values of momentum transfer $-t$, evolved from the model scale of $\mu_0^2=0.34$ GeV$^2$~\cite{Lan:2021wok} to the higher scale of $\mu^2=16$ GeV$^2$, shown as functions of the longitudinal momentum fraction $x$ carried by the parton.}
    \label{fig:xgpd-evolved}
    \end{center}
  \end{figure}

Furthermore, we perform QCD evolution of the pion’s unpolarized GPD, \( H(x, 0, t) \), to a higher scale \( \mu^2 \) using the next-to-next-to-leading order (NNLO) DGLAP equations~\cite{Altarelli:1977zs, Curci:1980uw, Furmanski:1980cm, Moch:2004pa, Vogt:2004mw}. For the numerical solution of these equations, we use the Higher Order Perturbative Parton Evolution Toolkit (HOPPET)~\cite{Salam:2008qg}. The evolution is carried out from the model scale \( \mu_0^2 = 0.34~\mathrm{GeV}^2 \)~\cite{Lan:2021wok} to \( \mu^2 = 16~\mathrm{GeV}^2 \), and the evolved results are presented in Fig.~\ref{fig:xgpd-evolved} for various values of the momentum transfer \( -t \).


At the model scale, the input includes only valence quark and gluon contributions, while the sea quarks are dynamically generated through the DGLAP evolution. This reflects the perturbative QCD mechanism by which gluon splitting populates the sea quark sector at higher scales.

As we probe the pion at short distance, and the momentum transfer \( -t \) increases, the quark GPDs decrease in magnitude and their peaks shift toward larger values of \( x \).
In the case of the evolved gluon GPD, we observe a pronounced emergence at small \( x \), indicating that gluons increasingly dominate the low-\( x \) region as the scale evolves.

\subsection{Impact Parameter-dependent GPDs}
The impact parameter (${\vec {b}}_\perp$) dependent GPDs describe the parton's probability density in the transverse position space when the momentum transfer in the longitudinal direction is zero ($\xi=0$). The impact parameter dependent GPDs are expressed by performing the Fourier transform of the GPDs with respect to the transerse momentum transfer ($\vec{\Delta}_\perp$) in the transverse direction~\cite{Burkardt:2002hr}
\begin{align}
&{\mathcal H}^{q,g}_{\pi}(x,0,{ \vec{b}}_\perp)= \int \frac{\mathrm{d}^2 \vec{ q}_{\perp}}{(2\pi)^2} e^{i{ \vec{\Delta}}_{\perp}\cdot \vec{ b}_{\perp}} H^{q,g}_\pi(x,0,-{\vec{\Delta}}^2_{\perp}).\label{eqn:ipd}
\end{align}

\begin{figure}[hbt!]
     \centering
     \begin{subfigure}[b]{0.47\textwidth}
         \centering
         \includegraphics[width=\textwidth]{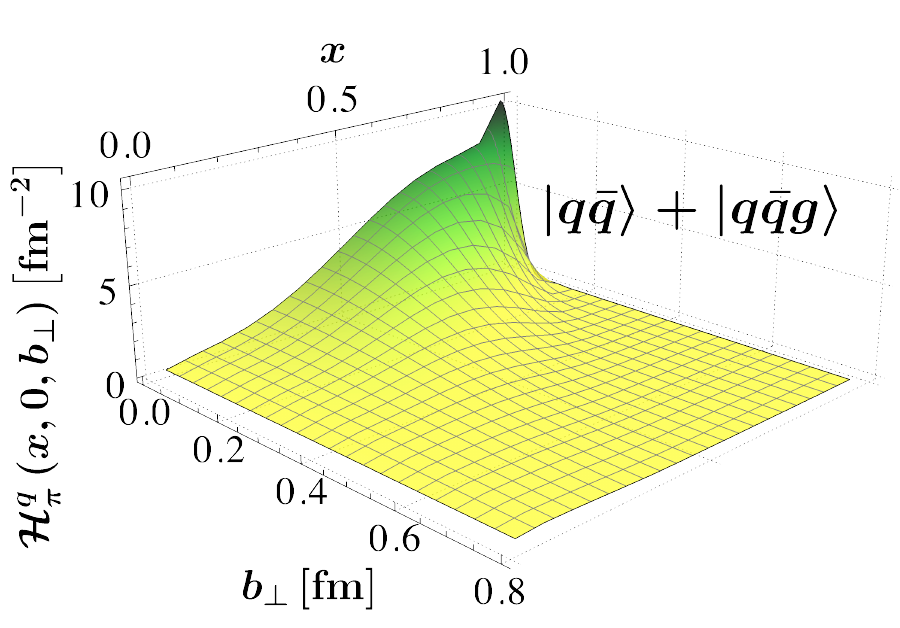}
         \caption{}
         \label{fig:ipdgpd-total}
     \end{subfigure} 
     \\
     \begin{subfigure}[b]{0.47\textwidth}
         \centering
         \includegraphics[width=\textwidth]{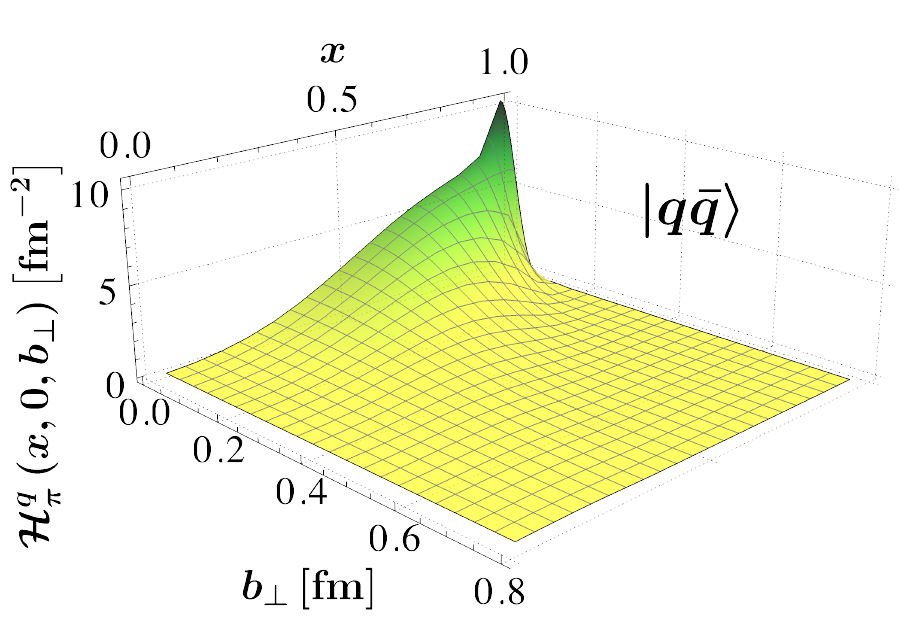}
         \caption{}
         \label{fig:ipdgpd-q-Fock1}
     \end{subfigure}
     \hfill
     \begin{subfigure}[b]{0.47\textwidth}
         \centering
         \includegraphics[width=\textwidth]{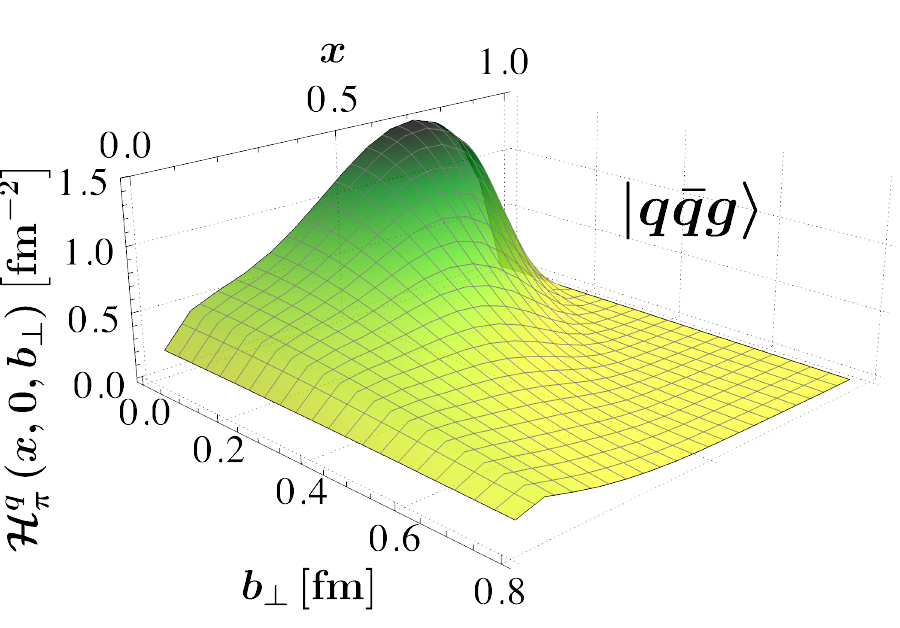}
         \caption{}
         \label{fig:ipdgpd-q-Fock22}
     \end{subfigure}
             \caption{
             The impact parameter dependent GPD of quark in the Fock sectors (a) $|q\bar{q}\rangle+|q\bar{q}g \rangle$, (b) $|q\bar{q}\rangle$, and (c) $|q \bar{q}g\rangle$ with respect to $x$ and $-t$.}
        \label{fig:ipdgpd-q-Fock2}
\end{figure}

\begin{figure}
    \centering
    \includegraphics[width=0.5\linewidth]{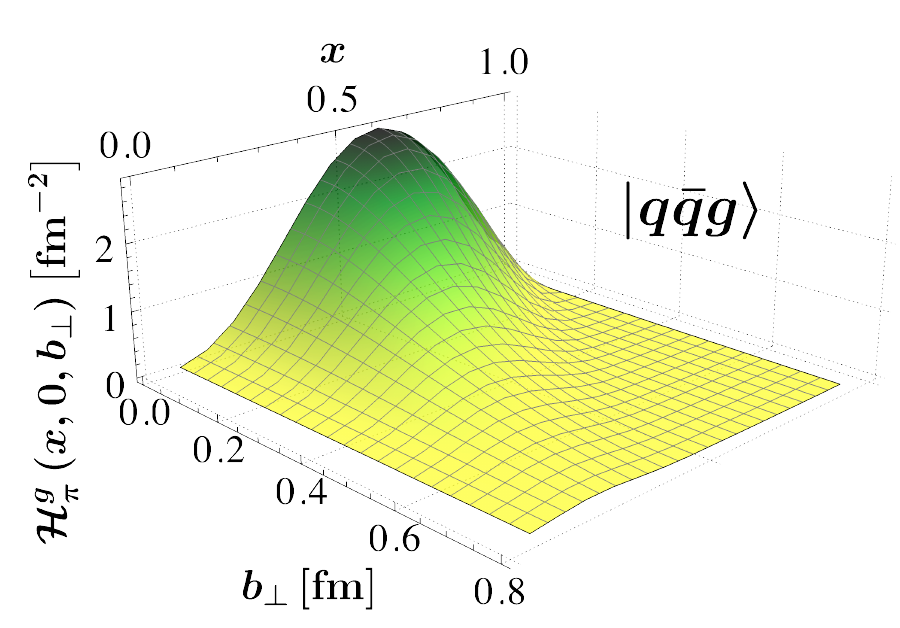}
    \caption{
    The impact parameter dependent GPD of gluon with respect to $x$ and $-t$.}
    \label{fig:ipdgpd-g-Fock2}
\end{figure}

In Figs.~\ref{fig:ipdgpd-q-Fock2} and \ref{fig:ipdgpd-g-Fock2}, we present the impact parameter dependent GPDs $\mathcal{H}(x,0,{\vec{b}}_\perp)$ for the quark and gluon in the pion, respectively. The distribution exhibits sharp peaks at the center of the pion $|{\vec{ b}}_\perp|=0$ when the quark carries large longitudinal momentum for the pion. 
An important feature is the decreasing width of the transverse distribution with increasing $x$, indicating that partons become more localized near the center of the meson as their longitudinal momentum increases.
This localization in transverse position space is consistent with the broadening of the GPDs in momentum space with increasing $x$, particularly in the $-t$ dependence, as shown in Figs.~\ref{fig:gpd-total} and~\ref{fig:gpd-g-Fock2}. 
This property of our pion's GPDs is also observed in other theoretical studies for the mesons~\cite{Kaur:2018ewq,Adhikari:2021jrh,Zhang:2021mtn,Raya:2021zrz,Albino:2022gzs,Kaur:2025gyr} as well as for the nucleon~\cite{Chakrabarti:2013gra,Mondal:2015uha,Maji:2017ill,Liu:2022fvl} inferring this feature is model-independent.
We also observe that the quark distribution arising from the leading Fock component is more localized in transverse position space than that from the next-to-leading Fock component, as shown in Figs.~\ref{fig:ipdgpd-q-Fock1} and \ref{fig:ipdgpd-q-Fock22}, respectively. This behavior can be attributed to their contrasting momentum-space characteristics: the contribution from the \( q\bar{q}g \) sector falls off much more rapidly than that of the \( q\bar{q} \) sector as shown in Fig.~\ref{fig:gpd-q-Fock1} and Fig.~\ref{fig:gpd-q-Fock2}, respectively.

The \( x \)-dependent squared transverse radius for the quark and gluon densities in the pion, in terms of the impact parameter dependent GPDs, can be defined as~\cite{Dupre:2016mai}
\begin{align}
    \langle {b}_\perp^2 \rangle^{q,g}(x) = \frac{\int \mathrm{d}^2 \vec{b}_\perp \, {b}_\perp^2 \, \mathcal{H}^{q,g}_\pi(x, 0, \vec{b}_\perp)}{\int \mathrm{d}^2 \vec{b}_\perp \, \mathcal{H}^{q,g}_\pi(x, 0, \vec{b}_\perp)},
\label{eqn:ipd_b2x}
\end{align}
and the total squared transverse radius, integrated over the parton momentum fraction, is expressed in terms of the PDFs as
\begin{align}
    \langle b^2_{\perp}\rangle^{q,g} &= \frac{\int \mathrm{d}x \mathrm{d}^2 \vec{b}_{\perp} b_{\perp}^2 \mathcal{H}_{\pi}^{q,g} (x,0,\vec{b}_\perp) }{\int \mathrm{d}x \mathrm{d}^2 \vec{b}_{\perp}  \mathcal{H}_{\pi}^{q,g} (x,0,\vec{b}_\perp) }.
\end{align}

\begin{figure}[hbt!]
     \centering
     \begin{subfigure}[b]{0.47\textwidth}
         \centering
         \includegraphics[width=\textwidth]{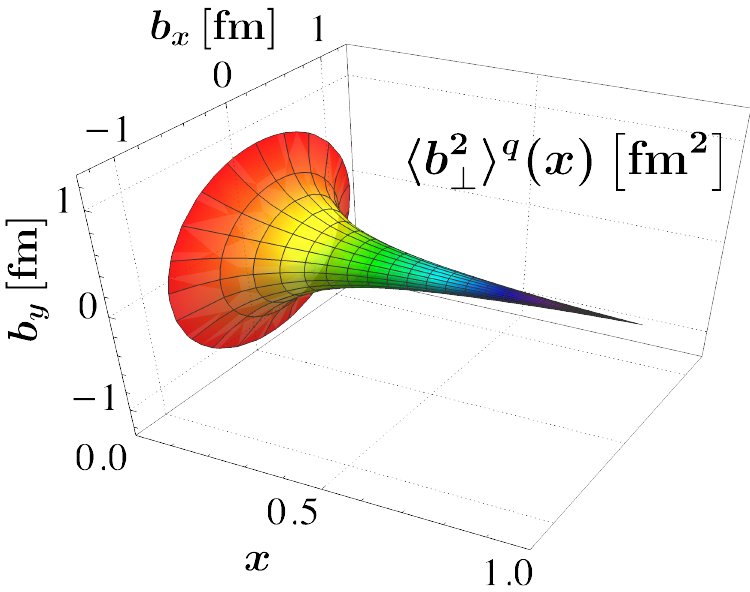}
         \caption{}
         \label{fig:b-space-q}
     \end{subfigure}
     \begin{subfigure}[b]{0.47\textwidth}
         \centering
         \includegraphics[width=\textwidth]{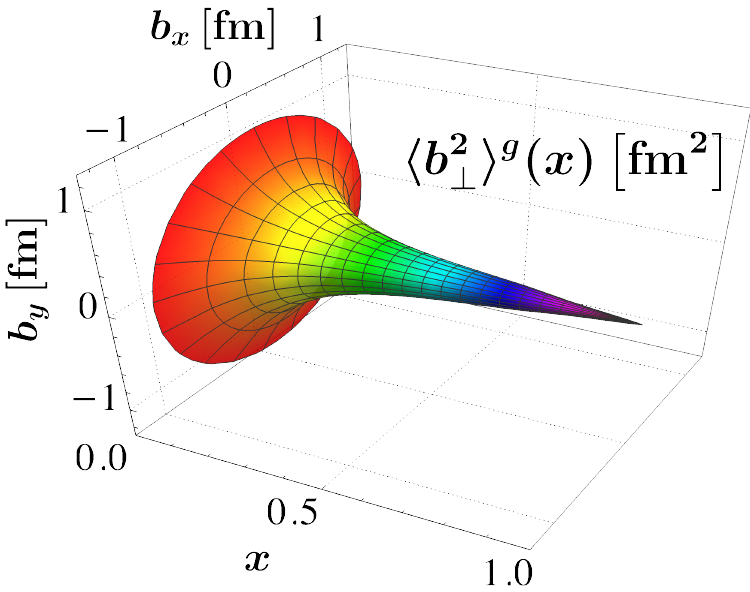}
         \caption{}
         \label{fig:b-space-g}
     \end{subfigure}
     \hfill
     \begin{subfigure}[b]{0.47\textwidth}
         \centering
         \includegraphics[width=\textwidth]{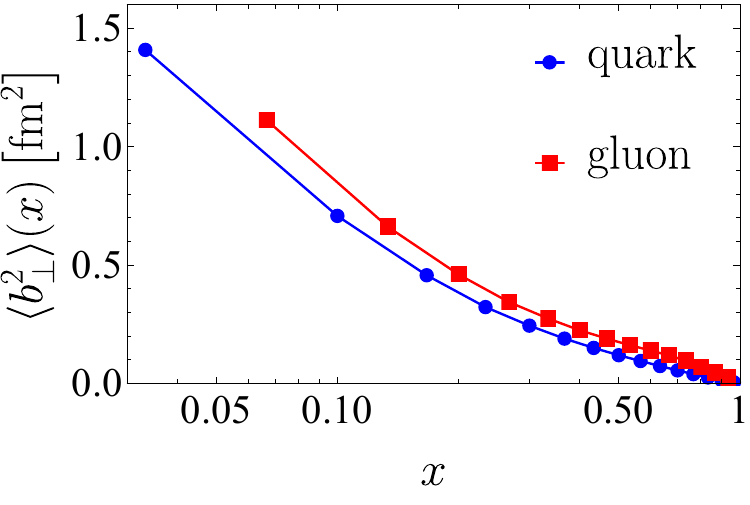}
         \caption{}
         \label{fig:bsquared}
     \end{subfigure}
             \caption{The $x$-dependent squared transverse radius of quark and gluon inside pion.}
        \label{fig:transverse-b}
\end{figure}

Figure~\ref{fig:transverse-b} displays the \( x \)-dependent squared transverse radius \( \langle {b}_\perp^2 \rangle^{q,g}(x) \), which characterizes the spatial extent of quark and gluon distributions in the transverse plane. 
As the longitudinal momentum fraction \( x \) decreases, the transverse size increases, a behavior consistent with the picture of partons becoming more delocalized at small \( x \)~\cite{Burkardt:2002hr}. 
This trend reflects the underlying dynamics of confinement, where low-momentum partons probe larger spatial regions of the pion. Notably, gluons exhibit a slightly broader transverse profile than quarks at a given \( x \), indicating a stronger spatial spread.

This behavior provides valuable insight into the spatial structure of the pion. 
The gluon distributions extend over a wider transverse area than quarks, meaning that from a large-distance perspective (or equivalently, at low transverse resolution), the gluon distribution forms a ``cloud", while the quark remain more concentrated near the center.

The integrated squared transverse radii, obtained by averaging over all \( x \) (weighted by corresponding PDFs~\cite{Dupre:2016mai}), quantify this difference in spatial extent: \( \langle {b}_\perp^2 \rangle^q \approx 0.35 \, \text{fm}^2 \) for the quark and \( \langle {b}_\perp^2 \rangle^g \approx 0.26 \, \text{fm}^2 \) for the gluon. 





\section{Conclusion and Outlook}
In this work, we used a light-front framework that includes both the leading (valence) and a higher Fock sector with one dynamical gluon to study the internal structure of the pion.
This study extends our earlier work on the pion~\cite{Lan:2021wok}, where we had focused on electromagnetic form factors and parton distribution functions. The present analysis builds on that foundation to provide a more comprehensive understanding of the pion’s internal dynamics, incorporating spatial and momentum distributions for both quark and gluon constituents.

We have computed key observables that offer a multidimensional view of the pion’s structure when the quark and gluon are unpolarized. Generalized Parton Distributions (GPDs) enabled us to probe the spatial distribution of partons in both longitudinal and transverse directions, while Transverse Momentum Dependent distributions (TMDs) revealed the correlation between transverse momentum and longitudinal momentum fractions. From our analysis of GPDs and TMDs, we observe that the inclusion of the higher Fock sector modifies the quark distributions, enriching their transverse position and momentum structures. Additionally, the 
$x$-dependent transverse radius shows that the gluon distribution is spatially broader than that of the quarks, indicating a more delocalized gluonic component within the pion.

This study offers a step forward in understanding how the pion is built from its partonic components. In future work, we plan to improve the model by including more gluon and sea quark contributions by expanding the Fock space of the pion. We also aim to compare our results with upcoming experimental data and lattice QCD predictions, which will help further refine our understanding of the pion. As a future goal, we plan to develop a more complete treatment of chiral symmetry breaking~\cite{Li:2022mlg}, thereby providing a deeper theoretical foundation for our framework.

\section*{Acknowledgements}
We thank Jiatong Wu for useful discussions. 
J. L. is supported by the Special Research Assistant Funding Project, Chinese Academy of Sciences, by the Gansu International Collaboration and Talents Recruitment Base of Particle Physics (2023-2027), by the Senior Scientist Program funded by Gansu Province, Grant No. 25RCKA008, by the National Natural Science Foundation of China under Grant No. 12305095, and the Natural Science Foundation of Gansu Province, China, Grant No. 23JRRA631.
S. K. is supported by Research Fund for International Young
Scientists, Grant No. 12250410251, from the National Natural Science Foundation of China (NSFC), and China Postdoctoral Science
Foundation (CPSF), Grant No. E339951SR0.
%
%
C. M. is supported by new faculty start up funding the Institute of Modern Physics, Chinese Academy of Sciences, Grants No. E129952YR0. 
X. Z. is supported by new faculty startup funding by the Institute of Modern Physics, Chinese Academy of Sciences, by Key Re- search Program of Frontier Sciences, Chinese Academy of Sciences, Grant No. ZDBS-LY-7020, by the Natural Science Foundation of Gansu Province, China, Grant No. 20JR10RA067, by the Foundation for Key Talents of Gansu Province, by the Central Funds Guiding the Local Science and Technology Development of Gansu Province, Grant No. 22ZY1QA006, by international partnership program of the Chinese Academy of Sciences, Grant No. 016GJHZ2022103FN, by the Strategic Priority Research Program of the Chinese Academy of Sciences, Grant No. XDB34000000, and by the National Natural Science Foundation of China under Grant No.12375143.
J. P. V. is supported by the Department of Energy under Grant No. DE-SC0023692.  A portion of the computational resources were also provided by Taiyuan Advanced Computing Center.

\bibliographystyle{apsrev}
\bibliography{ref}

 \end{document}